\documentclass{article}
\usepackage{spconf,amsmath,graphicx,hyperref}
\usepackage{booktabs}
\usepackage{multirow}
\usepackage[table]{xcolor}

\title{Prediction-Layer Branch Calibration for Multimodal Sentiment Analysis}

\name{Yulin Sun, Kele Xu\sthanks{Corresponding author.}, Yong Dou}
\address{National University of Defense Technology}

\begin{document}
%
\maketitle
\begin{abstract}
Multimodal sentiment analysis integrates textual, acoustic and visual cues, yet current language-model-based fusion methods typically leave prediction-layer branch allocation implicit. We introduce Branch-Calibrated Multimodal Language Fusion (BC-MLF), which explicitly models prediction-layer branch allocation through a Branch-Calibrated Task Head (BCHead), complemented by Fusion Token Contrastive Learning (FTCL) for sentiment-aware fusion-token regularization. FTCL organizes mean-pooled fusion-token representations according to continuous sentiment affinity, while BCHead combines fusion, text and audiovisual predictions through a lightweight sample-adaptive constrained mixture. Without modifying the fusion backbone, BC-MLF consistently improves the reproduced DeepMLF baseline and achieves the strongest results among the compared methods on CMU-MOSEI and CH-SIMS across classification and regression metrics. The controlled ablations show that sample-adaptive prediction-layer branch allocation consistently outperforms static branch aggregation. 
Code is available at \url{https://github.com/sunyulin0421/BC-MLF}.
\end{abstract}
\begin{keywords}
Multimodal Sentiment Analysis, Multimodal Language Fusion, Branch Allocation Calibration, Contrastive Learning, Sentiment Regression
\end{keywords}
\section{Introduction}
\label{sec:intro}

\begin{figure*}[t]
    \centering
    \includegraphics[width=\textwidth]{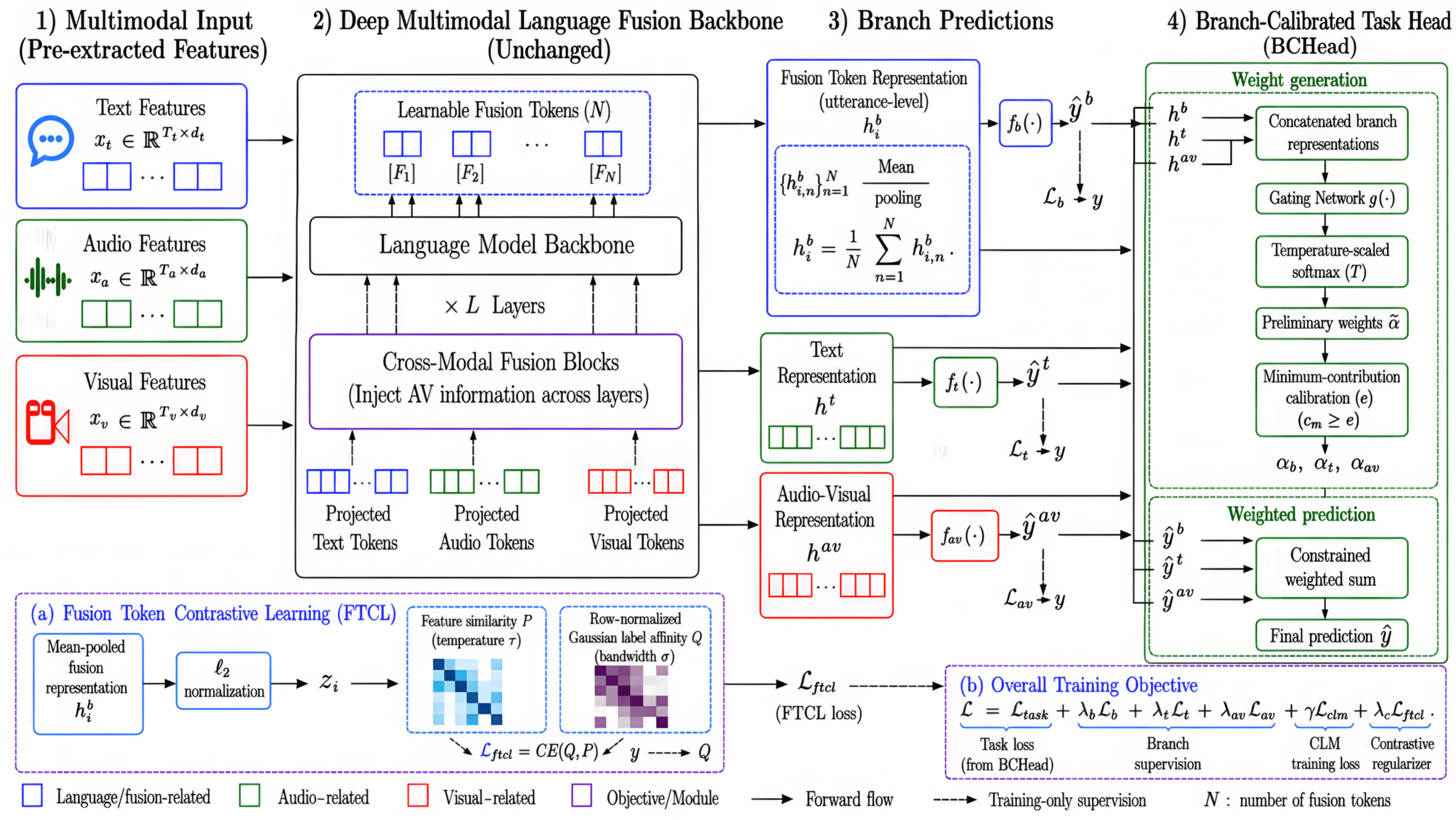}
    \caption{Branch-Calibrated Multimodal Language Fusion (BC-MLF). The unchanged fusion backbone produces fusion-token, text and audiovisual representations. During training, FTCL regularizes fusion-token geometry according to continuous sentiment affinity, whereas BCHead calibrates prediction-layer branch allocation through a lightweight constrained mixture.}
    \label{fig:framework}
\end{figure*}

Multimodal sentiment analysis (MSA) leverages heterogeneous language, acoustic, and visual signals to estimate human sentiment~\cite{baltrusaitis2019multimodal}. Early fusion architectures explored tensor and low-rank interactions~\cite{zadeh2017tfn,liu2018efficient,hou2019deep}, while later recurrent, hierarchical, and transformer-based models improved cross-modal representation learning~\cite{zadeh2018memory,georgiou2019deep,tsai2019MulT,hazarika2020misa,zhang2023almt}. Recent deep multimodal language fusion methods further introduce learnable fusion tokens to inject audiovisual information across language-model layers~\cite{georgiou2025deepmlf}.

Feature-level routing adapts multimodal fusion during representation construction~\cite{tsai2020multimodal}, while mutual-information and unified-learning objectives improve multimodal representation learning and supervision~\cite{han2021improving,hu2022unimse}. Regression-aware contrastive learning exploits continuous target structure~\cite{zha2023rank}, and prototype-guided methods address modality imbalance through representation- and optimization-level calibration~\cite{he2026pase}. These approaches primarily operate before the final decision stage, leaving the allocation of multiple supervised branch predictions at the decision stage implicit.

We address this prediction-layer allocation gap with Branch-Calibrated Multimodal Language Fusion (BC-MLF). The Branch-Calibrated Task Head (BCHead) converts fusion, text and audiovisual branch predictions into a sample-adaptive, minimum-constrained mixture, enabling explicit prediction-layer allocation at the decision stage. Complementarily, Fusion Token Contrastive Learning (FTCL) applies a Gaussian label-affinity objective to organize mean-pooled fusion-token representations according to continuous sentiment affinity. This representation-to-decision design jointly enhances sentiment-aware fusion geometry and enables sample-adaptive prediction-layer allocation, and we evaluate it on the English CMU-MOSEI~\cite{zadeh2016mosei} and Chinese CH-SIMS~\cite{yu2020sims} benchmarks.

The main contributions are as follows:
\begin{itemize}
    \item We formulate prediction-layer branch allocation in multimodal language fusion as an explicit calibration problem.
    \item We propose BC-MLF, which introduces a sample-adaptive, minimum-constrained mixture of existing branch predictions and complements it with Gaussian-affinity regularization of mean-pooled fusion-token representations.
    \item BC-MLF achieves the strongest results among the compared methods on CMU-MOSEI and CH-SIMS across classification and regression metrics, with controlled ablations verifying the advantage of sample-adaptive branch allocation beyond static aggregation.
\end{itemize}

\begin{table*}[t]
\centering
\caption{\textit{Comparison with representative state-of-the-art MSA methods on CMU-MOSEI and CH-SIMS.} $\dagger$: results reported in~\cite{georgiou2025deepmlf}; $*$: results reproduced; $\uparrow/\downarrow$: higher/lower is better. Bold: best result in each column. Results of \textbf{BC-MLF} are averaged over two random seeds.}
\label{tab:main_results}
\setlength{\tabcolsep}{8pt}
\begin{tabular}{l cccccc c cccc}
\toprule
\multirow{2}{*}{MODEL} & \multicolumn{6}{c}{CMU-MOSEI} & & \multicolumn{4}{c}{CH-SIMS} \\
\cmidrule(lr){2-7} \cmidrule(lr){9-12}
& Acc2$\uparrow$ & F1$\uparrow$ & MAE$\downarrow$ & Corr$\uparrow$ & Acc5$\uparrow$ & Acc7$\uparrow$ & & Acc2$\uparrow$ & F1$\uparrow$ & MAE$\downarrow$ & Corr$\uparrow$ \\
\midrule
LF-DNN$\dagger$        & 82.78 & 82.38 & 0.558 & 0.731 & --    & --    & & 76.68 & 76.48 & 0.446 & 0.567 \\
TFN$\dagger$           & 82.23 & 81.47 & 0.573 & 0.718 & --    & --    & & 77.07 & 76.94 & 0.437 & 0.582 \\
MAG-BERT$\dagger$      & 84.87 & 84.85 & 0.539 & 0.764 & --    & --    & & 74.44 & 71.75 & 0.492 & 0.399 \\
MulT$\dagger$          & 84.07 & 83.93 & 0.564 & 0.731 & 53.97 & 52.56 & & 78.56 & 78.66 & 0.453 & 0.564 \\
MISA$\dagger$          & 84.51 & 84.47 & 0.549 & 0.759 & 53.57 & 51.96 & & 76.54 & 76.59 & 0.447 & 0.563 \\
TETFN$\dagger$         & 85.20 & 85.18 & 0.544 & 0.759 & 55.54 & 53.74 & & 79.21 & 79.05 & 0.419 & 0.592 \\
Self-MM$\dagger$       & 84.26 & 84.24 & 0.532 & 0.765 & 55.52 & 53.85 & & 80.04 & 80.44 & 0.425 & 0.595 \\
ALMT$\dagger$          & 85.23 & 85.32 & 0.539 & 0.766 & 54.64 & 53.05 & & 78.16 & 78.16 & 0.433 & 0.575 \\
DeepMLF$*$             & 86.08 & 86.10 & 0.505 & 0.802 & 57.58 & 55.91 & & 81.95 & 82.36 & 0.373 & 0.706 \\
\midrule
\rowcolor{blue!8}
\textbf{BC-MLF}          & \textbf{87.50} & \textbf{87.50} & \textbf{0.495} & \textbf{0.808} & \textbf{58.36} & \textbf{56.69} & & \textbf{82.94} & \textbf{83.12} & \textbf{0.363} & \textbf{0.729} \\
\bottomrule
\end{tabular}
\end{table*}

\definecolor{groupcolor}{gray}{0.92}
\definecolor{bestcolor}{rgb}{0.9,0.95,1.0}

\begin{table}[!t]
\centering
\caption{Ablation studies of BC-MLF. The results demonstrate the effectiveness of FTCL and BCHead, and show that sample-adaptive prediction-layer allocation provides additional benefits beyond static branch aggregation.}
\normalsize
\setlength{\tabcolsep}{7pt}
\begin{tabular}{l c c c c}
\toprule
\textbf{Method} &
\textbf{Acc2$\uparrow$} &
\textbf{F1$\uparrow$} &
\textbf{MAE$\downarrow$} &
\textbf{Corr$\uparrow$} \\
\midrule

\rowcolor{groupcolor}
\multicolumn{5}{c}{\textbf{MOSEI}} \\
\midrule

DeepMLF & 86.08 & 86.10 & 0.505 & 0.802 \\
+ FTCL & 86.75 & 86.74 & 0.501 & 0.805 \\
+ BCHead & 87.40 & 87.39 & 0.495 & 0.807 \\
+ FTCL + Uniform & 86.18 & 86.13 & 0.518 & 0.798 \\
+ FTCL + Global & 86.15 & 86.14 & 0.514 & 0.799 \\
BC-MLF ($\epsilon=0$) & 87.19 & 87.17 & 0.496 & 0.807 \\

\rowcolor{bestcolor}
\textbf{BC-MLF} &
\textbf{87.50} &
\textbf{87.50} &
\textbf{0.495} &
\textbf{0.808} \\

\midrule

\rowcolor{groupcolor}
\multicolumn{5}{c}{\textbf{CH-SIMS}} \\
\midrule

DeepMLF & 81.95 & 82.36 & 0.373 & 0.706 \\
+ FTCL & 82.17 & 82.52 & 0.370 & 0.710 \\
+ BCHead & 82.82 & 82.93 & 0.368 & 0.719 \\
+ FTCL + Uniform & 82.06 & 82.26 & 0.376 & 0.709 \\
+ FTCL + Global & 82.16 & 82.37 & 0.376 & 0.710 \\
BC-MLF ($\epsilon=0$) & 82.85 & 82.93 & 0.366 & 0.722 \\

\rowcolor{bestcolor}
\textbf{BC-MLF} &
\textbf{82.94} &
\textbf{83.12} &
\textbf{0.363} &
\textbf{0.729} \\

\bottomrule
\end{tabular}
\label{tab:ablation}
\end{table}

\section{Methodology}
\label{sec:method}

\subsection{Overview}

Branch-Calibrated Multimodal Language Fusion (BC-MLF) builds on a multimodal language fusion backbone with learnable fusion tokens. For utterance $i$, the backbone produces $N$ fusion-token states $\{h_{i,n}^{b}\}_{n=1}^{N}$, together with a text representation $h_i^{t}$ and an audiovisual representation $h_i^{av}$. We mean-pool the fusion-token states to obtain an utterance-level representation:
\begin{equation}
    h_i^{b}
    =
    \frac{1}{N}
    \sum_{n=1}^{N} h_{i,n}^{b}.
\end{equation}
We denote $h_i^{b}$ as the fusion-token representation in the experiments. The three branch predictors then produce
\begin{equation}
    \hat{y}_i^{b}=f_b(h_i^{b}), \quad
    \hat{y}_i^{t}=f_t(h_i^{t}), \quad
    \hat{y}_i^{av}=f_{av}(h_i^{av}).
\end{equation}
A standard deep-fusion task head predicts sentiment from the concatenated representations:
\begin{equation}
    \hat{y}_i^{task}
    =
    f_{task}([h_i^{b};h_i^{t};h_i^{av}]).
\end{equation}
This formulation provides a joint prediction but leaves branch-level allocation implicit. As shown in Fig.~\ref{fig:framework}, BC-MLF addresses this gap by coupling FTCL, which structures the mean-pooled fusion-token space, with BCHead, which calibrates the mixture of branch predictions for the final decision.

\subsection{Fusion Token Contrastive Learning}

Continuous sentiment labels provide graded supervision for pairwise sample relations. FTCL transfers this structure to the utterance-level, mean-pooled fusion-token representation $h_i^{b}$ through a soft contrastive objective. For a mini-batch of $B$ samples, each representation is $\ell_2$-normalized as $z_i=h_i^{b}/\lVert h_i^{b}\rVert_2$.

For each non-self pair $i\neq j$, we define the label affinity and its row-normalized target distribution as
\begin{equation}
    s_{ij}
    =
    \exp\left(
        -\frac{(y_i-y_j)^2}{2\sigma^2}
    \right),
    \qquad
    q_{ij}
    =
    \frac{s_{ij}}
    {\sum_{k\neq i}s_{ik}},
\end{equation}
where $y_i$ denotes the sentiment label and $\sigma$ controls the affinity bandwidth. The corresponding representation-similarity distribution is
\begin{equation}
    p_{ij}
    =
    \frac{\exp(z_i^\top z_j/\tau)}
    {\sum_{k\neq i}\exp(z_i^\top z_k/\tau)},
    \qquad i\neq j,
\end{equation}
where $\tau$ is the contrastive temperature. FTCL minimizes the cross-entropy between the label-affinity and representation-similarity distributions:
\begin{equation}
    \mathcal{L}_{ftcl}
    =
    -\frac{1}{B}
    \sum_{i=1}^{B}
    \sum_{j\neq i}
    q_{ij}\log p_{ij}.
\end{equation}
FTCL uses a soft Gaussian affinity distribution to regularize pooled fusion-token representations according to continuous sentiment affinity. Applied only during training, it introduces no additional inference-time module.

\subsection{Branch-Calibrated Task Head}

FTCL structures sentiment-aware fusion-token geometry, while BCHead uses the three branch representations to predict sample-adaptive allocation logits for the final decision:
\begin{equation}
    u=[h^{b};h^{t};h^{av}],
    \qquad
    r=g(u).
\end{equation}
A temperature-scaled softmax converts the logits into a raw allocation distribution:
\begin{equation}
    \tilde{\alpha}
    =
    \mathrm{softmax}(r/T),
\end{equation}
where $T$ controls its sharpness. BCHead maps the raw distribution to a simplex with a minimum allocation weight $\epsilon$:
\begin{equation}
    \alpha_m
    =
    (1-3\epsilon)\tilde{\alpha}_m+\epsilon,
    \quad m\in\{b,t,av\}.
\end{equation}
For $0\leq\epsilon\leq1/3$, the calibrated weights satisfy $\alpha_m\geq\epsilon$ and $\sum_m\alpha_m=1$. The final prediction is
\begin{equation}
    \hat{y}
    =
    \alpha_b\hat{y}^{b}
    +\alpha_t\hat{y}^{t}
    +\alpha_{av}\hat{y}^{av}.
\end{equation}

\subsection{Training Objective}

BC-MLF optimizes the original deep-fusion objectives together with BCHead-based task supervision and FTCL regularization. The complete objective is
\begin{equation}
\begin{aligned}
    \mathcal{L}
    ={}& \mathcal{L}_{task}(\hat{y},y)
    +\lambda_b\mathcal{L}_{b}
    +\lambda_t\mathcal{L}_{t}
    +\lambda_{av}\mathcal{L}_{av} \\
    &+\gamma\mathcal{L}_{clm}
    +\lambda_c\mathcal{L}_{ftcl},
\end{aligned}
\end{equation}
where \(L_b,L_t,L_{av}\) denote the losses of the fusion, text, and audiovisual predictors, respectively. $\mathcal{L}_{clm}$ denotes the causal language-modelling loss retained from the backbone. The coefficients $\lambda_b$, $\lambda_t$, $\lambda_{av}$, $\gamma$ and $\lambda_c$ weight the objectives. During inference, the auxiliary objectives are inactive; prediction uses the learned branch predictors and BCHead.

\section{Experiments}
\label{sec:experiments}

\subsection{Experimental Setup}

We evaluate BC-MLF under the DeepMLF feature and evaluation protocol on CMU-MOSEI and CH-SIMS~\cite{zadeh2016mosei,yu2020sims}, enabling controlled comparison with the deep multimodal language fusion baseline. Acc-2, F1, MAE and Corr are reported on both datasets, with Acc-5 and Acc-7 additionally reported on MOSEI. Acc-2 and F1 follow the Non0 protocol. Results are averaged over two random seeds, 1990 and 1991.

Hyperparameters of FTCL and BCHead are determined on the validation set and fixed across experiments. All other training settings follow DeepMLF~\cite{georgiou2025deepmlf}. BCHead introduces only $0.166$M and $0.100$M additional trainable parameters on CMU-MOSEI and CH-SIMS, respectively, while FTCL adds no parameters and is inactive during inference.

\subsection{Main Results}

Table~\ref{tab:main_results} compares BC-MLF with representative MSA methods. BC-MLF obtains the strongest result on every reported metric across CMU-MOSEI and CH-SIMS. Relative to DeepMLF, it improves Acc-2, F1 and Corr while reducing MAE, highlighting the benefit of explicit prediction-layer branch allocation for deep multimodal fusion.

\subsection{Ablation Study}

Table~\ref{tab:ablation} evaluates the contributions of FTCL, BCHead and adaptive allocation against static aggregation controls. Adding FTCL improves every reported metric over DeepMLF on both benchmarks, while BCHead provides the larger gain. Uniform and Global controls further show that the gain comes from sample-adaptive allocation rather than static branch aggregation. The full model further improves on the unconstrained variant ($\epsilon=0$) across all four metrics on both datasets. These results establish complementary roles: FTCL structures fusion-token geometry, BCHead performs sample-adaptive branch allocation beyond static mixtures, and the allocation floor provides an additional refinement of the prediction mixture.

\subsection{Diagnostic Analysis}

\begin{figure}[t]
    \centering
    \includegraphics[width=0.80\columnwidth]{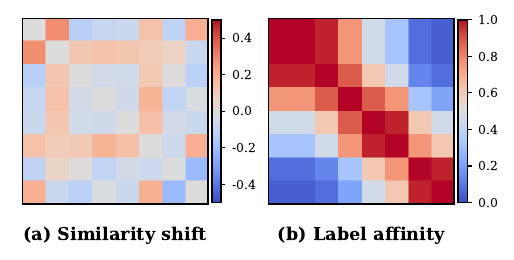}
    \caption{Pairwise similarity shifts in mean-pooled fusion-token representations between DeepMLF and BC-MLF, shown alongside the continuous label-affinity reference used by FTCL on CH-SIMS.}
    \label{fig:ftcl_shift_affinity}
\end{figure}

\begin{figure}[t]
    \centering
    \includegraphics[width=\columnwidth]{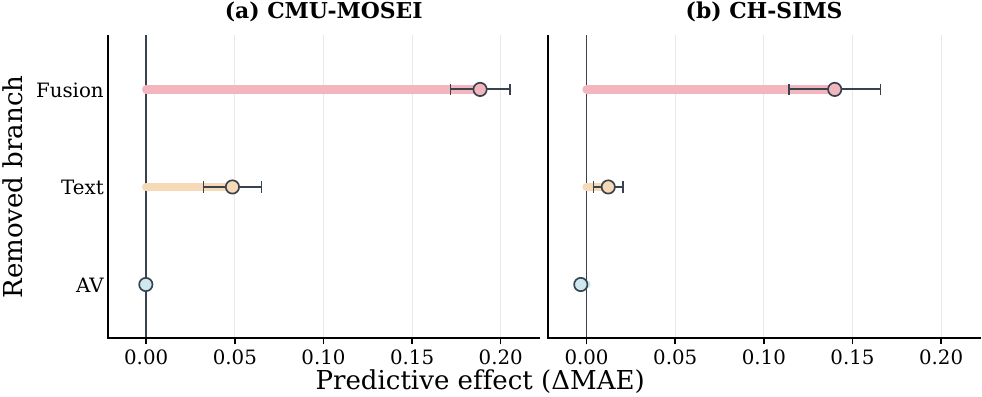}
    \caption{Frozen-gate branch-removal analysis on CMU-MOSEI and CH-SIMS. Mixture weights are fixed from the complete input while each branch prediction is removed, and $\Delta\mathrm{MAE}$ quantifies the conditional predictive effect. Points denote the mean over two seeds with error bars indicating standard deviation.}
    \label{fig:branch_effect}
\end{figure}



Figure~\ref{fig:ftcl_shift_affinity} visualizes fusion-token similarity shifts, while Table~\ref{tab:ftcl_alignment} quantifies their alignment with continuous sentiment affinity. FTCL improves Spearman correlation and reduces $D_{\mathrm{KL}}(Q\|P)$ on both benchmarks, indicating more sentiment-aware fusion-token geometry.

For branch $m$, we remove the corresponding branch prediction while keeping the mixture weights computed from the complete input fixed. The predictive effect is measured as $\Delta_m=\mathrm{MAE}_{-m}-\mathrm{MAE}_{full}$. As shown in Fig.~\ref{fig:branch_effect}, removing the fusion prediction produces the largest increase in MAE on both benchmarks, followed by the text prediction, whereas the audiovisual prediction has little effect under the learned allocation. This analysis reveals the conditional predictive effects captured by the sample-adaptive branch allocation, with fusion predictions providing the dominant predictive effect on both benchmarks.

\begin{table}[t]
\centering
\caption{Fusion-token alignment with continuous sentiment affinity
(mean $\pm$ standard deviation over two seeds).}
\label{tab:ftcl_alignment}
\small
\setlength{\tabcolsep}{2.5pt}
\renewcommand{\arraystretch}{1.05}
\begin{tabular}{@{}llcc@{}}
\toprule
Dataset & Model
& Spearman $\uparrow$
& $D_{\mathrm{KL}}(Q\|P)\downarrow$ \\
\midrule
MOSEI
& DeepMLF
& 0.3701 $\pm$ 0.0022
& 0.2410 $\pm$ 0.0025 \\
& DeepMLF + FTCL
& \textbf{0.3862 $\pm$ 0.0043}
& \textbf{0.2358 $\pm$ 0.0021} \\
\midrule
CH-SIMS
& DeepMLF
& 0.2610 $\pm$ 0.0312
& 0.1793 $\pm$ 0.0035 \\
& DeepMLF + FTCL
& \textbf{0.2749 $\pm$ 0.0387}
& \textbf{0.1757 $\pm$ 0.0026} \\
\bottomrule
\end{tabular}
\end{table}

\section{Conclusion}
\label{sec:conclusion}

We introduced Branch-Calibrated Multimodal Language Fusion (BC-MLF), which explicitly models prediction-layer branch allocation through a lightweight, sample-adaptive mixture of fusion, text and audiovisual predictions. Without modifying the fusion backbone, BC-MLF achieved the strongest results among the compared methods across classification and regression metrics on CMU-MOSEI and CH-SIMS. These results show that sample-adaptive prediction-layer branch allocation provides an effective decision-stage complement to deep multimodal fusion beyond static branch aggregation.

\bibliographystyle{IEEEbib}
\bibliography{refs}

\end{document}